%% file: paper.tex
\documentclass[12pt]{article}
			\usepackage{ifthen}			
			\usepackage{graphicx}			
			\usepackage[sort&compress]{natbib}	
			\usepackage{abbrvate}			
			\usepackage{amsmath}			
			\usepackage{amssymb}			
			\newcommand{\bibpath}{bibtex}
			
			\newabbr{BSM}{beyond the standard model}
			\newabbr{CC}{charged current}
			\newabbr{CDMS}{cryogenic dark matter search}
			\newabbr{CKM}{Cabbibo-Kobayashi-Maskawa}
			\newabbr{CoGeNT}{coherent germanium neutrino technology}
			\newabbr{CRESST}{cryogenic rare event search with superconducting thermometers}
			\newabbr{DM}{dark matter}
			\newabbr{DAMA}{dark matter}
			\newabbr{EWSB}{electroweak symmetry breaking}
			\newabbr{EOM}{equation of motion}
			\newabbr{EM}{electromagnetic}
			\newabbr{KARMEN}{Karlsruhe-Rutherford medium energy neutrino}
			\newabbr{LEP}{large electron positron collider}
			\newabbr{LHC}{large hadron collider}
			\newabbr{LIBRA}{large sodium iodide bulk for rare processes}
			\newabbr{LSND}{liquid scintillator neutrino detector}
			\newabbr{MiniBooNE}{mini booster neutrino experiment}
			\newabbr{RGE}{renormalization group equation}
			\newabbr{SSB}{spontaneous symmetry breaking}
			\newabbr{SM}{standard model}
			\newabbr{SUSY}{supersymmetry}
			\newabbr{T2K}{Tokai to Kamioka}
			\newabbr{VEV}{vacuum expectation value}
			\newabbr{WIMP}{weakly interacting massive particle}
			\input{\cmdpath/cite_commands}

			\input{\cmdpath/math-calculus}
			\input{\cmdpath/math-algebra}
			\input{\cmdpath/math-fractions}
			\input{\cmdpath/math-parentheses}
			\input{\cmdpath/format-science-misc}

			\newcommand{\vwk}{v_\text{wk}}

\begin{document}
\thispagestyle{empty}

\hfill{\sc UG-FT-302/13}

\vspace*{-2mm}
\hfill{\sc CAFPE-172/13}

\vspace{25pt}
\begin{center}
{\textbf{\Large Contamination of Dark Matter Experiments\\
from Atmospheric Magnetic Dipoles}}

\vspace{25pt}

A.~Bueno, M.~Masip, P.~S\'anchez-Lucas, N.~Setzer
\vspace{12pt}

\textit{
CAFPE and Departamento de F{\'\i}sica Te\'orica y del Cosmos}\\ 
\textit{Universidad de Granada, 18071 Granada, Spain}\\
\vspace{15pt}
\texttt{abueno,masip,patriciasl,nsetzer@ugr.es}
\end{center}

\vspace{25pt}

\date{\today}

\abbrstyle{expandnone}
\begin{abstract}
Dark matter collisions with heavy nuclei (Xe, Ge, Si, Na) may produce recoils observable at direct-search experiments. Given that some of these experiments are yielding conflicting information, however, it is worth asking if physics other than dark matter may produce similar nuclear recoils.  We examine under what conditions an atmospherically-produced neutral particle with a relatively large magnetic dipole moment could fake a dark matter signal.  We argue that a very definite flux could explain the signals seen at \abbr{DAMA}/\abbr{LIBRA}, \abbr{CDMS}/Si and \abbr{CoGeNT} consistently with the bounds from XENON100 and \abbr{CDMS}/Ge. To explore the plausibility of this scenario, we discuss a concrete model with 10--50 MeV sterile neutrinos that was recently proposed to explain the \abbr{LSND} and \abbr{MiniBooNE} anomalies.
\end{abstract}


\newpage

	\pagestyle{plain}	
	\abbrstyle{plain}
	\abbrreset
	\abbrmakeused[CDMS]
	\abbrmakeused[CKM]
	\abbrmakeused[CoGeNT]
	\abbrmakeused[CRESST]
	\abbrmakeused[DAMA]
	\abbrmakeused[LIBRA]
	\abbrmakeused[KARMEN]
	\abbrmakeused[LEP]
	\abbrmakeused[LHC]
	\abbrmakeused[LSND]
	\abbrmakeused[MiniBooNE]
	\abbrmakeused[T2K]
	\abbrmakeused[WIMP]

\section{Introduction}

The current tune of \abbr{DM} experiments is that of discordance: \abbr{DAMA}/\abbr{LIBRA} \cite{Bernabei:2008yi,Bernabei:2010mq}, \abbr{CoGeNT} \cite{Aalseth:2010vx}, \abbr{CRESST}-II \cite{Angloher:2011uu}, and \abbr{CDMS}/Si \cite{Agnese:2013rvf} report signals while XENON100 and the rest of experiments see nothing.  In addition, the signal seen by \abbr{DAMA}/\abbr{LIBRA} is seemingly inconsistent with the ones at \abbr{CoGeNT} and \abbr{CDMS} (at least when assuming a \abbr{WIMP} as \abbr{DM}).  Given this cacophonous state, it seems reasonable to consider alternative explanations for the current experimental data.  In particular, the fact that the \abbr{DAMA}/\abbr{LIBRA} modulation closely follows\footnote{Assuming a simple $A\cos[\omega(t-t_0)]$ dependence, one obtains $t_0 = 185\pm 15$ days for the muon flux \cite{Selvi:2010fk} and $t_0 = 146\pm 7$ days in \abbr{DAMA}/\abbr{LIBRA} \cite{Bernabei:2010mq} indicating an incompatibility of a common phase by 2 standard deviations.  A further investigation of the correlation is given in \cite{Chang:2011eb}. We think that a dedicated study of the correlation between atmospheric parameters (such as temperature and pressure) and the \abbr{DAMA}/\abbr{LIBRA} signal could provide evidence of an atmospheric origin if this is indeed the case.} that of the muon flux \cite{Nygren:2011xu} suggests that it may be atmospheric in origin. This idea is not new, as a number of authors have proposed that the two phenomena are related either directly \cite{Blum:2011jf,Ralston:2010bd} or indirectly through neutron backgrounds \cite{Nygren:2011xu}.  In this work we take an orthogonal approach and consider the possibility that these observations may be a genuine sign of new physics, just \emph{not} \abbr{DM}.  Thus, we address the question of whether  direct search experiments exclusively probe \abbr{DM} or whether they could  be revealing a {\it contamination} by some unexpected physics.

\begin{figure}[!ht]
\begin{center}
\includegraphics[width=0.498\linewidth]{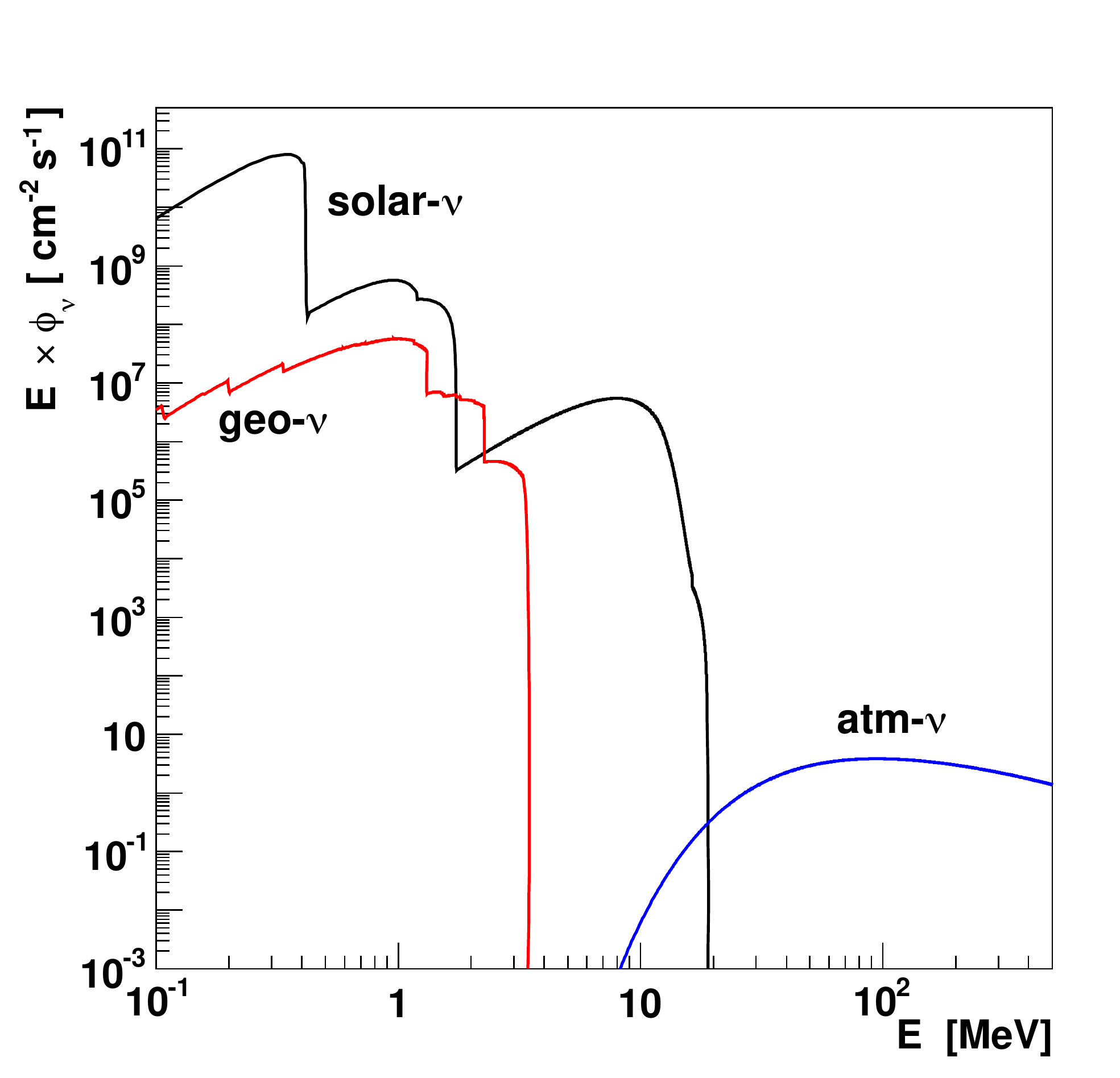}
\includegraphics[width=0.46\linewidth]{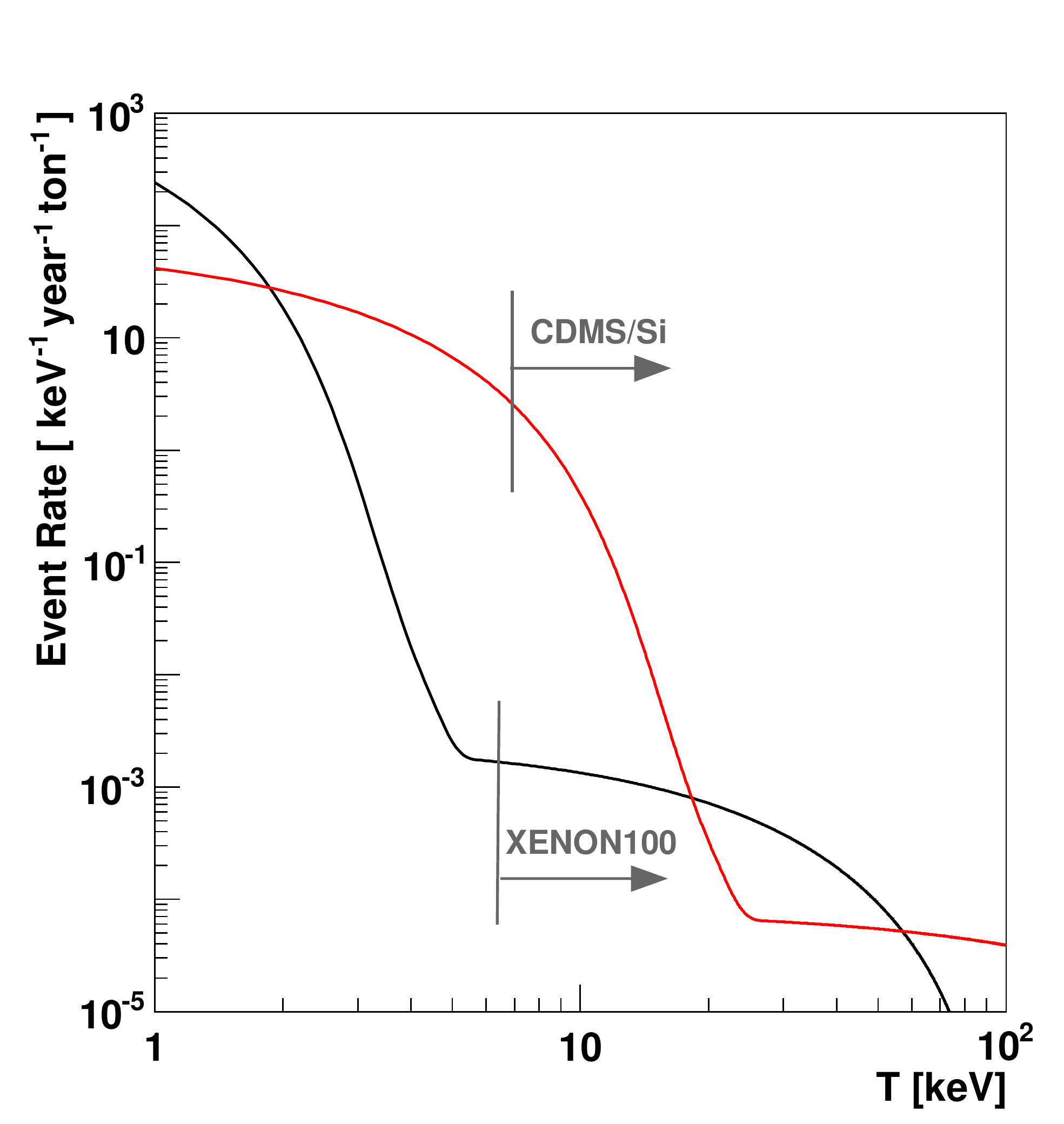} 
\end{center}
\caption{To the left is shown the ambient neutrino fluxes; to the right the corresponding recoil distribution that they produce at XENON100 \cite{Aprile:2012nq} and \abbr{CDMS}/Si \cite{Agnese:2013rvf}, with the energy threshold of each experiment also indicated.  The recoil distribution has a ``kink'' which represents the transition from solar neutrinos to atmospheric neutrinos (the geoneutrino contribution is not significant).  For the complete details on obtaining these plots see \cite{Sanchez-Lucas:2012fk}.} 
\label{Fig:flux.nu.and.recoils}
\end{figure}

To obtain a likely new physics candidate, one starting point is to consider standard background sources and modify them in order to increase or tune their effects. The usual set-up for a direct-search experiment is to have some target mass and look for recoils off a nucleus. Therefore, in these experiments, ambient neutrinos that interact coherently with the nucleus via $Z$-exchange represent one such irreducible background. In \fig{flux.nu.and.recoils}(right) we plot our estimate \cite{Sanchez-Lucas:2012fk} for this neutrino background at XENON100 \cite{Aprile:2012nq} and \abbr{CDMS}/Si \cite{Agnese:2013rvf} (whose lighter target implies larger recoils $T$) for the neutrino flux given in \fig{flux.nu.and.recoils}(left).  At energies below $20$ MeV the ambient flux is dominated by solar neutrinos whereas at higher energies neutrinos come from the decay of atmospheric muons and mesons. At $E\approx 2$ MeV there is a significant contribution to the flux from geoneutrinos produced by radiactive ${}^{238}$U, ${}^{235}$Th and ${}^{40}$K \cite{Araki:2005qa}, but this contribution is not important in direct-search experiments (\fig{flux.nu.and.recoils}, right).
Solar neutrinos produce $\mathcal{O}(100)$ events per year and ton of $^{131}$Xe, but all of them are below threshold for XENON100.  In \abbr{CDMS}/Si solar neutrinos yield $\mathcal{O}(100)$ events per year and ton, but the experimental threshold only admits $\mathcal{O}(10)$ of these.  The contribution of atmospheric neutrinos at $10^{-3}$/yr/ton in XENON100 and $10^{-4}$/yr/ton at \abbr{CDMS} is far too low to be significant; however, this need not be the case if there are other neutrino species that are produced frequently in muon decays.

The above then suggests a way to introduce new physics: a sterile neutrino or, generically, a new neutral particle produced either atmospherically or in the sun.  Sterile neutrinos of solar origin have been explored previously \cite{Harnik:2012ni,Pospelov:2012gm}, so we will focus on atmospheric production.  We provide the details and properties of such a hypothetical, atmospherically-produced particle in \Sec{neutral.particle.properties}.  In \Sec{dark.matter.contamination}, we discuss how this particle appears at various \abbr{DM} experiments.  In \Sec{heavy.neutrino} we introduce a model that serves as a concrete example of this particle.  Finally, we conclude our discussion in \Sec{conclusion}.

\section{Magnets from the Atmosphere}
\label{Sec:neutral.particle.properties}

Starting from the idea of an additional neutrino species, we contemplate the properties that a hypothetical particle should have to generate a signal at \abbr{DM} experiments. A likely candidate would possess the following properties:
\begin{itemize}
\item{neutral charge}
\item{produced in muon decays (directly or within a decay chain) with frequency of $1$ particle per $100$-$1000$ decays}
\item{mass in excess of $10$ MeV}
\item{lifetime between $10^{-6}$ s and $1$ s}
\item{large magnetic dipole moment}
\end{itemize}
The requirement of neutral charge is evident for mimicking a \abbr{DM} signal.  The second property links the atmospheric production to muons, so that the hypothetical particle's flux may modulate just as the muon flux---thus providing the possibility to explain annual modulations.  The condition on its mass permits the particle to avoid astrophysical constraints (for example production in the sun or other stars), while the lifetime range affords the possibility of being atmospherically produced and still reaching the detector without introducing cosmological problems (such as disrupting big-bang nucleosynthesis).  The final condition of a large magnetic dipole moment provides a means of giving the particle a long-range \abbr{EM} interaction without introducing another gauge symmetry.

An \abbr{EM} dipole will introduce photon-mediated interactions with nuclei that are rather different from the standard neutrino interactions and thus allow for a significant contamination in \abbr{DM} experiments.  Elastic neutrino interactions are mediated by a $Z$ boson and as such are short-distance processes.  If we ignore the nuclear form factor, such interactions tend to give nuclear recoils, $T$, right below
\begin{equation}
T_\text{max}	= \frac{ 2M \inp{E^2 - m^2} }{ M^2 + 2 M E + m^2 },
\label{Eq:Def.Tmax}
\end{equation}
with $E$ the energy of the neutrino, $m$ its mass, and $M$ the mass of the nucleus. Furthermore, the form factor will suppress elastic collisions with recoils having a momentum transfer $q^2 = 2 M T \gtrsim \inp{100 \text{ MeV}}^2$; therefore,  
\begin{equation}
T'_\text{max}	= \frac{ (0.1\text{ GeV})^2 }{ 2 M }.
\label{Eq:Def.TmaxPrime}
\end{equation}
These $Z$-mediated processes will then be dominated by recoils near $\min\{ T_\text{max},T'_\text{max}\}$, with lower recoils giving a non-significant contribution to the total cross section.  In contrast, the photon-mediated interaction of a dipole includes all recoils below $\min \{ T_\text{max},T'_\text{max}\}$:~all of them contribute equally\footnote{The atomic form factor will suppress the cross section also at very low recoils.} to the total cross section.

We therefore introduce our hypothetical particle as a Dirac fermion, $\Psi$, with a magnetic dipole moment $\mu$.  Its lagrangian is then
\begin{equation}
\mathcal{L} \supset \half \mu \overline \Psi \sigma_{\alpha\beta} \Psi F^{\alpha\beta} + m \overline \Psi \Psi\,,
\label{Eq:lagrangian.dm.contaminator}
\end{equation}
where $F^{\mu\nu}$ is the field strength for the photon.  With the langragian of \eq{lagrangian.dm.contaminator}, we may now discuss how this particle, a sterile Dirac neutrino, could be seen at direct-search \abbr{DM} experiments.

\section{Contaminating Dark Matter Experiments}
\label{Sec:dark.matter.contamination}

With the lagrangian given in \eq{lagrangian.dm.contaminator}, our particle will collide with nuclei through an \abbr{EM} dipole interaction:~assuming a spin-$1/2$ target of mass $M$ and charge $Z$, the differential cross-section for the elastic scattering reads
\begin{equation}
T \, \deriv{\sigma}{T} 	= \alpha Z^2 \mu^2  F_Z^2\inp{q^2} \inp{ 1 - \frac{T}{2 M} \, \frac{2 E M - m^2}{E^2 - m^2} },
\label{Eq:T.dsigmadT}
\end{equation}
where $\alpha$ is the fine-structure constant and we take the form factor \cite{Gninenko:1998nn}
\begin{align}
F_Z^2(q^2)
	& = 
		\begin{cases}
		\displaystyle \frac{a^4 q^4}{ \inp{1 + a^2 q^2}^2 }	& q^2 \le t_0 = 7.39 m_e^2	\\
		\displaystyle \frac{1}{ \inp{1 + q^2/d}^2 }		& q^2 > t_0
		\end{cases}
\label{Eq:Form.Factor}
\end{align}
with $a = 111.7\text{ MeV}^{-1} Z^{-1/3} / m_e$ and $d = 0.164 A^{-2/3}\text{ GeV}^2$.  As we mentioned, the form factor suppresses elastic collisions of $q^2 = 2 M T \gtrsim (100 \text{ MeV})^2$ since at these energies inelastic processes dominate.  

\begin{figure}[!pht]
\begin{center}
\includegraphics[width=0.48\linewidth]{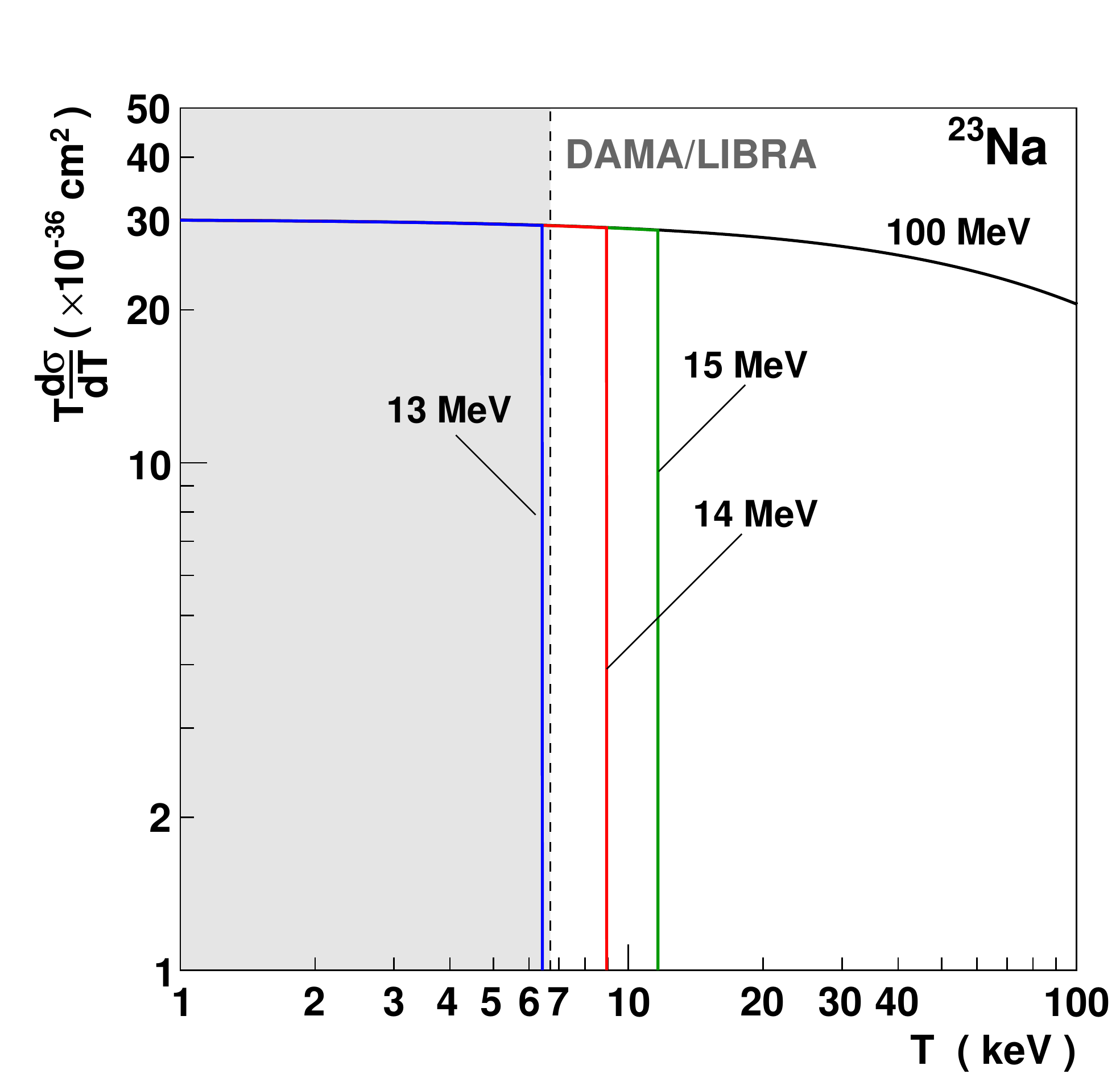}
\includegraphics[width=0.48\linewidth]{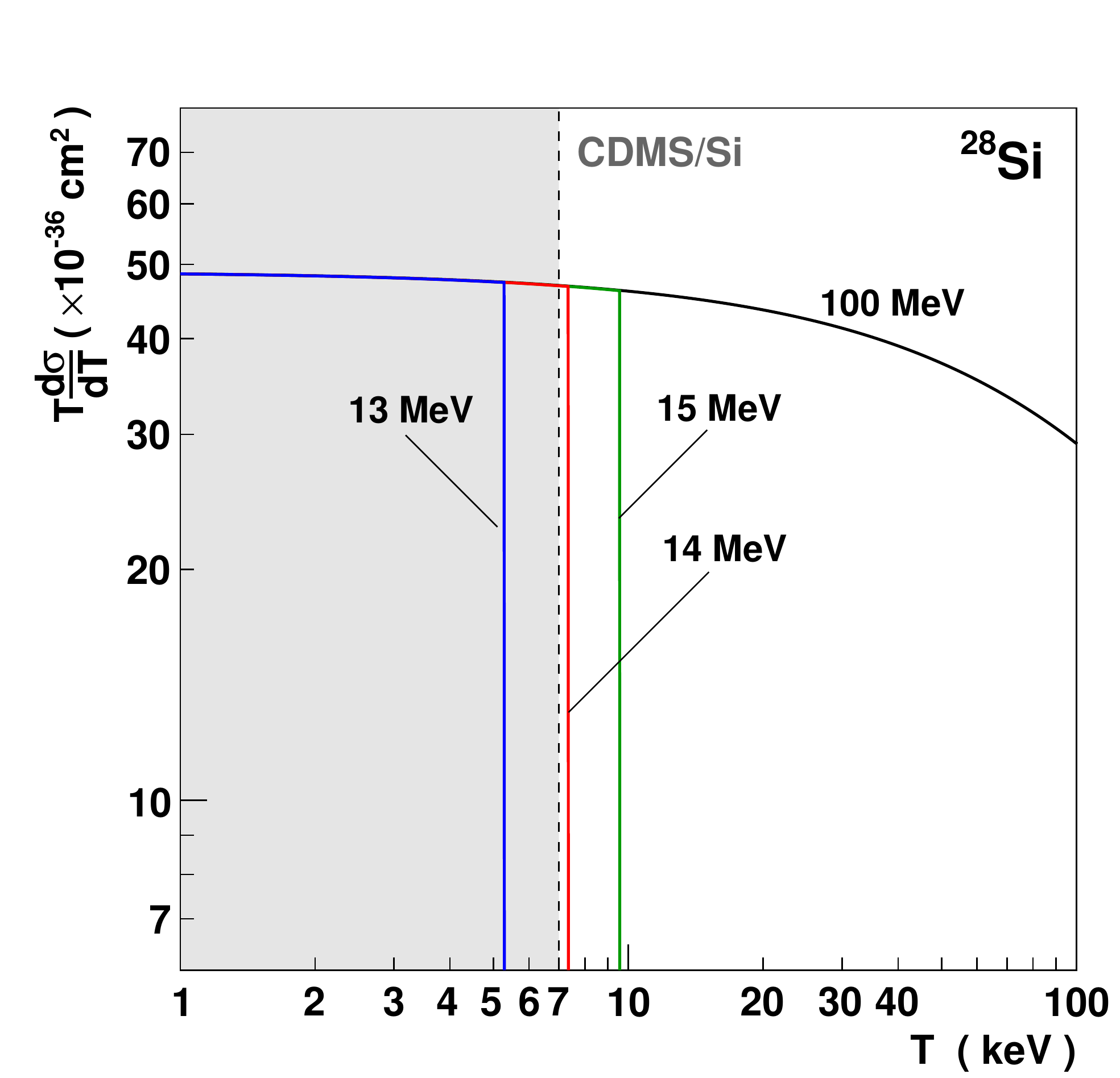}
\includegraphics[width=0.48\linewidth]{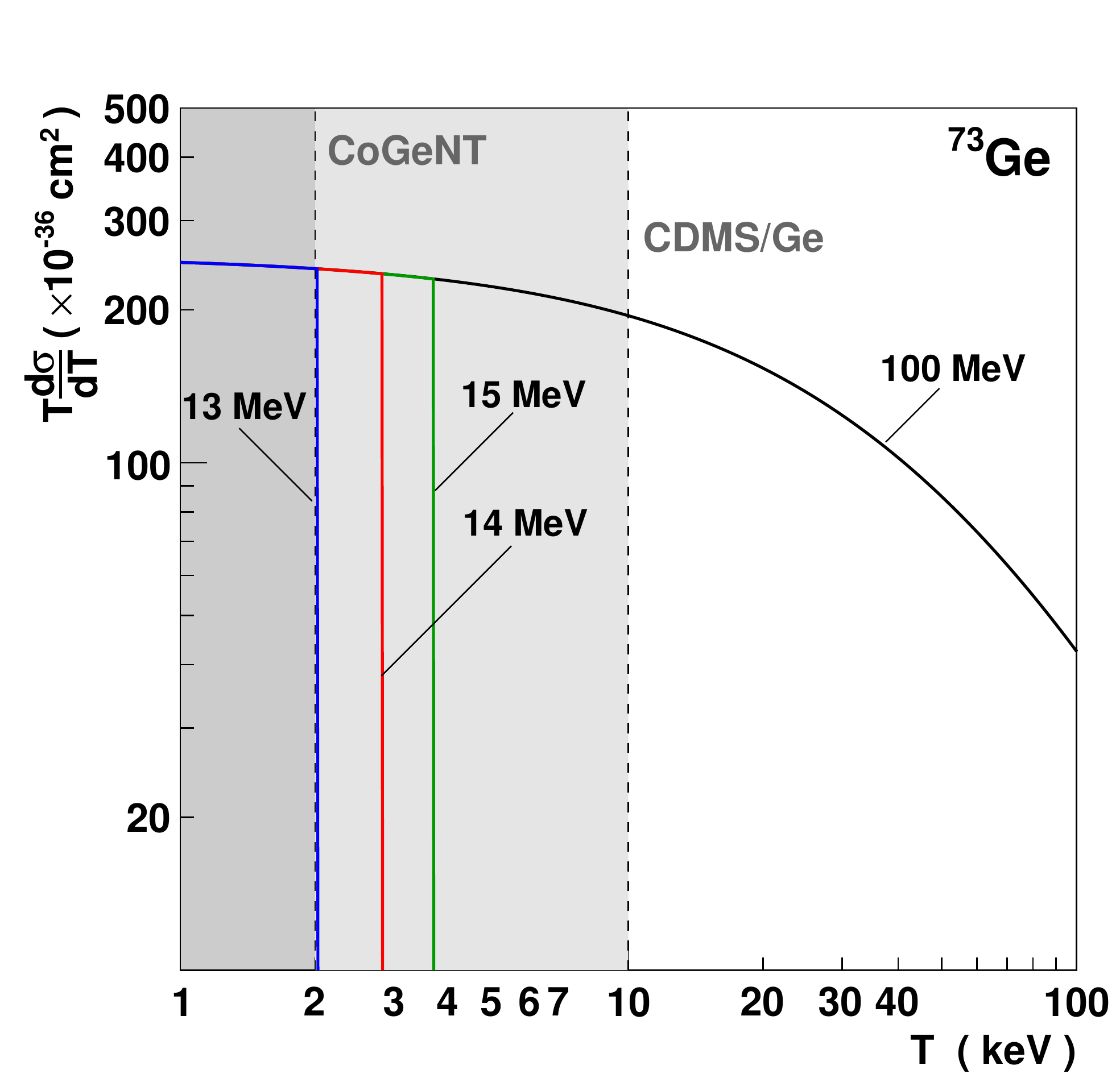}
\includegraphics[width=0.48\linewidth]{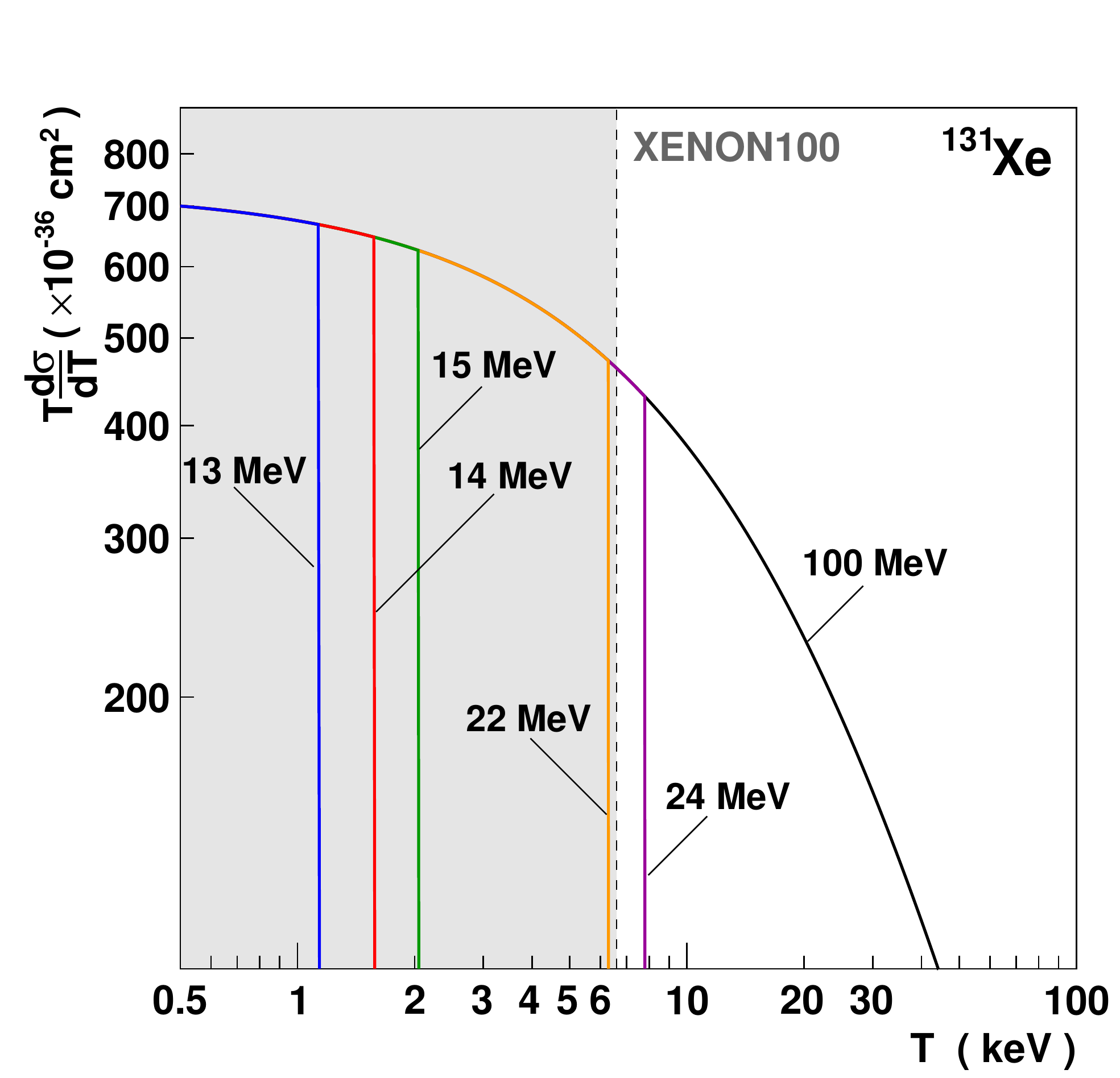} 
\end{center}
\caption{The recoil times the differential cross section versus the recoil of an elastic collision of our hypothetical particle with four different target nuclei.  The shaded regions denote recoils  below threshold for the indicated experiments.  The mass of the particle is taken to be $m = 10 \text{ MeV}$ and its magnetic moment is $\mu = 10^{-6} \mu_B$.  Note that a particle with an energy between $13$ and $14$ MeV can be seen by \abbr{DAMA}/\abbr{LIBRA} and \abbr{CoGeNT}, but not \abbr{CDMS} or XENON100.  For an energy between $14$ and $15$ MeV the particle can be seen by \abbr{CDMS}/Si, but not XENON100; it is not until the energy of the particle is between $22$ and $24$ MeV that it can be seen by XENON100.}
\label{Fig:T.dsigmadT.vs.T} 
\end{figure}

The differential cross section is plotted in \fig{T.dsigmadT.vs.T} for four target elements and several incident energies.  We have assumed that the particle has a mass of $m = 10 \text{ MeV}$ and a magnetic dipole moment $\mu = 10^{-6} \mu_B$ (we show later that this magnetic moment is the largest obtainable in the natural embedding of this setup).
As can be seen in \fig{T.dsigmadT.vs.T}, at large energies ($E \sim 100\text{ MeV}$) the maximum recoil is determined solely by the nuclear form factor; at low energies ($E \sim \mathcal{O}(10)\text{ MeV}$), however, kinematics force the differential cross section to zero before the form factor becomes relevant.  It is this kinematic feature that can be used to explain the various experiments. 

Given that $T\deriv{\sigma}{T} \approx \text{const}$, an incident particle of energy $E$ will interact at a given detector with a \textit{visible} cross section (\textit{i.e.}, $T \ge T_\text{thres})$ of
\begin{equation}
\sigma(E) \approx  \alpha Z^2 \mu^2 \ln\frac{T_{\text{max}}(E)}{T_{\text{thres}}}
\label{Eq:cross.secttion}.
\end{equation}
As a consequence, for a given detector exposure, $X$, a flux of $N(E) = \Phi(E)\,\Delta E$ of such particles reaching the detector will produce 
\begin{equation}
N_{\text{events}} \approx \alpha Z^2 \mu^2 \frac{X}{m_{\text{mat}}} N(E)  
\ln\frac{T_{\text{max}}(E)}{T_{\text{thres}}}
\label{Eq:N.Events}
\end{equation}
where $\Phi(E) = dN/dA\,dt\,dE$ is the differential flux (per unit area, time, and energy) and $m_\text{mat}$ is the mass of the target material in the detector (either the atomic or molecular mass in kg to match the units of the exposure).  If we bin our flux in $1$ MeV increments, the plots of \fig{T.dsigmadT.vs.T} show that the energies between $13$--$14$ MeV introduce recoils that are seen both at \abbr{DAMA}/\abbr{LIBRA} and \abbr{CoGeNT}.  Taking the targets and exposures in \tbl{Experiment.Data} (and an average energy in the bin of $E = 13.5 \text{ MeV}$), we predict the ratio 
\begin{align}
\frac{N_{\text{events}}^{\text{\abbr{DAMA}/\abbr{LIBRA}}}}{N_{\text{events}}^{\text{\abbr{CoGeNT}}}}	& \sim 700 
\sim \frac{12\,400}{17}
\label{Eq:N.Events.DAMA.over.N.Events.CoGeNT}
\end{align}
for a mass $m = 10 \text{ MeV}$ and independent of the magnetic dipole $\mu$.  This should be compared to the values from the actual experiments, which are also given in \tbl{Experiment.Data}.
\begin{table}[!t]
\begin{center}
\begin{tabular}{|c|c|c|c|c|c|}
\hline
Experiment					& Threshold (keV)	& Exposure (kg d)	& Material	& Tgt.		& Num. Events			\\
\hline
\abbr{DAMA}/NaI\cite{Bernabei:2010mq}		& 6.7			& $109\,500$ 		& NaI(Tl)	& Na		& $\sim 4\,300$ 			\\
\abbr{DAMA}/\abbr{LIBRA}\cite{Bernabei:2010mq}	& 6.7			& $317\,697$ 		& NaI(Tl)	& Na		& $\sim 12\,400$ 		\\
\abbr{CoGeNT}\cite{Aalseth:2010vx}		& $2$			& $18.48$		& Ge		& Ge		& $\sim 7$--$13$			\\
\abbr{CDMS}/Si\cite{Agnese:2013rvf}		& $7$			& $140.2$		& Si		& Si		& $3$				\\
\abbr{CDMS}/Ge\cite{Ahmed:2009zw}		& $10$			& $612$			& Ge		& Ge		& $0$				\\
XENON100\cite{Aprile:2012nq}			& $6.6$			& $7636.4$		& Xe		& Xe		& $0$				\\
\hline
\end{tabular}
\end{center}
\caption{The experimental data (taken from papers indicated) and the target nuclei of interest.  Note that \abbr{CoGeNT} does not directly give the number of events; however, in Fig.~3 of \cite{Aalseth:2010vx} they plot the allowed number of events for a \abbr[>>]{DM} \abbr{WIMP} of mass $7$ and $10$ MeV.  It is from these curves that we estimate the allowed number of events shown above.}
\label{Table:Experiment.Data}
\end{table}
Note that the prediction of \eq{N.Events.DAMA.over.N.Events.CoGeNT} is fairly well-obeyed by the data.

While the $13$--$14$ MeV bin is seen by both \abbr{DAMA}/\abbr{LIBRA} and \abbr{CoGeNT}, the experiments \abbr{CDMS} and XENON100 only see higher-energy particles due to their heavier targets or higher thresholds.  For \abbr{CDMS}/Si, the $14$--$15$ MeV bin represents the lowest neutrino energies yielding detectable recoils.  If we take the number of \abbr{DAMA}/\abbr{LIBRA} events as the base of comparison, the ratio of events at \abbr{CDMS}/Si to \abbr{DAMA}/\abbr{LIBRA} is given by $N(14.5\text{ MeV})/ N(13.5\text{ MeV})$---a quantity fixed due to the $3$ events at \abbr{CDMS}/Si. With the assistance of \eq{N.Events}, this events ratio sets the ratio of the fluxes,
\begin{align}
\frac{ \Phi(14.5\text{ MeV}) }{ \Phi(13.5\text{ MeV}) }
	& \sim  0.045,
\label{Eq:Flux.CDMS.over.Flux.DAMA}
\end{align}
independently of $\mu$.  Similarly, the $22$--$24$ MeV bin is the first one producing nuclear recoils above the threshold at XENON100.  As XENON100 does not see any events, we obtain an upper bound on the ratio of fluxes,
\begin{align}
\frac{ \Phi(23.5\text{ MeV}) }{ \Phi(14.5\text{ MeV}) }
	& \lesssim  2.9\E{-4},
\label{Eq:Flux.XENON.over.Flux.CDMS}
\end{align}
which implies that XENON100 would see $0.1$ event (i.e., none) for every $3$ observed by \abbr{CDMS}/Si.  It should be emphasized that, given Eqs.~\eqn{Flux.CDMS.over.Flux.DAMA} and \eqn{Flux.XENON.over.Flux.CDMS}, neither \abbr{DAMA}/\abbr{LIBRA} nor \abbr{CoGeNT} would see any significant signal coming from  dipoles with energies in excess of $14$ MeV.  In fact, all experiments would only see a significant number of events near their energy thresholds.

It is possible to further constrain the flux if an exponential dependence on momentum\footnote{Typically the literature refers to a flux dependent on energy rather than momentum; however, this assumes that $E \gg m$ so that $\abs{\vec{p}} \approx E$.  Here we have $E \sim m$ so it is necessary to distinguish the two.} is assumed,
\begin{equation}
\Phi \propto \inp{E^2 - m^2}^{-\gamma}.
\label{Eq:Flux.E.dependence}
\end{equation}
With this ansatz, the ratio of \eq{Flux.CDMS.over.Flux.DAMA} implies the rather steep fall-off of $\gamma \approx 10$.  Taking the flux to obey this in the entire region of interest implies that
\begin{align}
\frac{  N_{\text{events}}^{\text{XENON100}}  }{  N_{\text{events}}^{\text{\abbr{CDMS}/Si}}  }	& \sim 7.4\E{-5}
\label{Eq:N.Events.XENON.over.N.Events.CDMS}
\end{align}
so that the ratio of events at XENON100 to \abbr{CDMS}/Si is much less than the bound of \eq{Flux.XENON.over.Flux.CDMS}---actually, XENON100 would need about $1000$ years of running time to get just one event.  While this rapid fall-off is consistent with current experiments, it seems difficult to reconcile with an atmospheric production in muon decays.  Since the details of how the spectrum may be shifted to lower energies are rather model-specific, we defer that discussion until the next section.  In any case, it is worth pointing out that the functional dependence of the flux is entirely unknown and would not be determined by the dark matter experiments.

\section{A Concrete Model: The Heavy Neutrino}
\label{Sec:heavy.neutrino}

Having established the generic properties that our dipole should have in order to consistently explain direct-search \abbr{DM} experiments, it is time to introduce an example model.  As previously mentioned, massive sterile (gauge singlet) neutrinos are a natural possibility: they can mix with the muon neutrino flavor and appear in a fraction of muon decays. In fact, although
the three-flavor picture fits the solar, atmospheric, and reactor data \cite{Conrad:2007ea} well, there has been a persistent anomaly in experiments with neutrino beams from particle accelerators. Both \abbr{LSND} \cite{Athanassopoulos:1996jb,Athanassopoulos:1996wc,Athanassopoulos:1997pv,Aguilar:2001ty} and \abbr{MiniBooNE} \cite{AguilarArevalo:2007it,AguilarArevalo:2008rc,AguilarArevalo:2009xn,AguilarArevalo:2010wv} have observed an excess of $3$ events with an electron in the final state per each 1000 \abbr{CC} $\nu_\mu$ interactions.  These excesses may be explained by the addition of two sterile neutrinos with masses around $1\text{ eV}$ \cite{Abazajian:2012ys} or in the $10$'s of MeV \cite{Gninenko:2010pr,Masip:2012ke}.  The model we introduce shall be based upon this latter interpretation, which is also able to interprete the negative result at \abbr{KARMEN} \cite{Armbruster:2002mp}.

Our basic set up will be defined by two Dirac neutrinos, $\nu_h$ and $\nu_{h^\prime}$, in addition to the three standard flavors (which we assume to be Majorana particles). We will use hereafter a two-component spinor notation, with $f$ and $f^c$ indicating fermion and antifermion fields both of left-handed chirality.  We have then
\begin{align}
\nu_h		& = \begin{pmatrix} N_h \\ \inp{N_h^c}^\dagger \end{pmatrix}				&
\nu_{h^\prime}	& = \begin{pmatrix} N_{h^\prime} \\ \inp{N_{h^\prime}^{c}}^\dagger \end{pmatrix}
\label{Eq:nu.h.to.two.component}.
\end{align}
We will assume that $N_h$ (the left-handed component of $\nu_h$) mixes with the muon-neutrino, $\nu_\mu$; the effective lagrangian should thus contain
\begin{equation}
- \mathcal{L}_\text{eff}	\supset \frac{y \vwk}{\sqrt{2}} \nu_\mu N_h^c + m_h N_h N_h^c + m_{\nu_\mu} \nu_\mu \nu_\mu + \text{h.c.}
\label{Eq:lagrangian.nuh.numu.mixing}
\end{equation}
The mass eigenstates are then given by
\begin{align} 
N_h^\prime	& = N_h \cos \theta  + \nu_{\mu} \sin \theta \,,
	&
\nu_\mu^\prime
	& = - N_h \sin \theta + \nu_{\mu} \cos \theta,
\label{Eq:mu.and.nuh.mass.eigenstates}
\end{align}
where (neglecting corrections of $\mathcal{O}\inp{m_{\nu_\mu}/m_h}$)
\begin{equation}
\sin \theta	= \frac{ y \vwk }{ \sqrt{2 m_h^2 + y^2 \vwk^2 / 2} } \approx U_{\mu h}
\label{Eq:nuh.numu.mixing.angle}.
\end{equation}
In order to reproduce the \abbr{LSND} anomaly \cite{Gninenko:2010pr} the model requires $m_h \approx 50\text{ MeV}$ and $|U_{\mu h}|^2 = 0.003$ (which is obtained for $y \vwk /\sqrt{2} \approx 2.7\text{ MeV}$).  In addition, explaining the \abbr{MiniBooNE} anomaly consistently with \abbr{T2K} data \cite{Abe:2011sj} requires the neutrino $\nu_h$ to decay with $99$\% branching ratio into another sterile neutrino \cite{Masip:2012ke}---namely $\nu_{h^\prime}$---through the process
\begin{equation}
\nu_h \to \nu_{h^\prime} \gamma
\label{Eq:nuh.decay.to.nuhprime.gamma}.
\end{equation}
This decay is induced by the operator
\begin{equation}
\mathcal{L}_\text{eff}
	\supset  \half \mu_\text{tr}^{h h^\prime} N_h^c \sigma _{\alpha\beta} N_{h^\prime} F^{\alpha\beta} + \text{h.c.}\,,
\end{equation}
which yields the rather short lifetime $\tau_h \approx 5\E{-9}\text{ s}$ for $\mu_\text{tr}^{h h^\prime} \approx 2.4\E{-8}\mu_B$.  Thus, as $\nu_h$ would be produced in $0.1$--$1$\% of all muon and kaon decays \cite{Masip:2011qb}, $\nu_{h'}$ would be very abundant in the atmosphere provided its lifetime is sufficiently long. 

The above supplies a plausible production mechanism for the $\nu_{h^\prime}$, but we still have to justify the relatively large value of the \abbr{EM} transitions and the magnetic moments. We will first show that the required transitions, $\mu^{i h}_\text{tr}$, can be naturally obtained in left-right \cite{Mohapatra:1991ng} extensions of the standard model (which provide a natural embedding for these sterile neutrinos), but that the simplest extension does not imply a large enough magnetic moment $\mu_{h^\prime}$. We will then discuss a possible scenario that could boost the value of this magnetic moment.

\begin{figure}[!t]
\begin{center}
\includegraphics[scale=0.7]{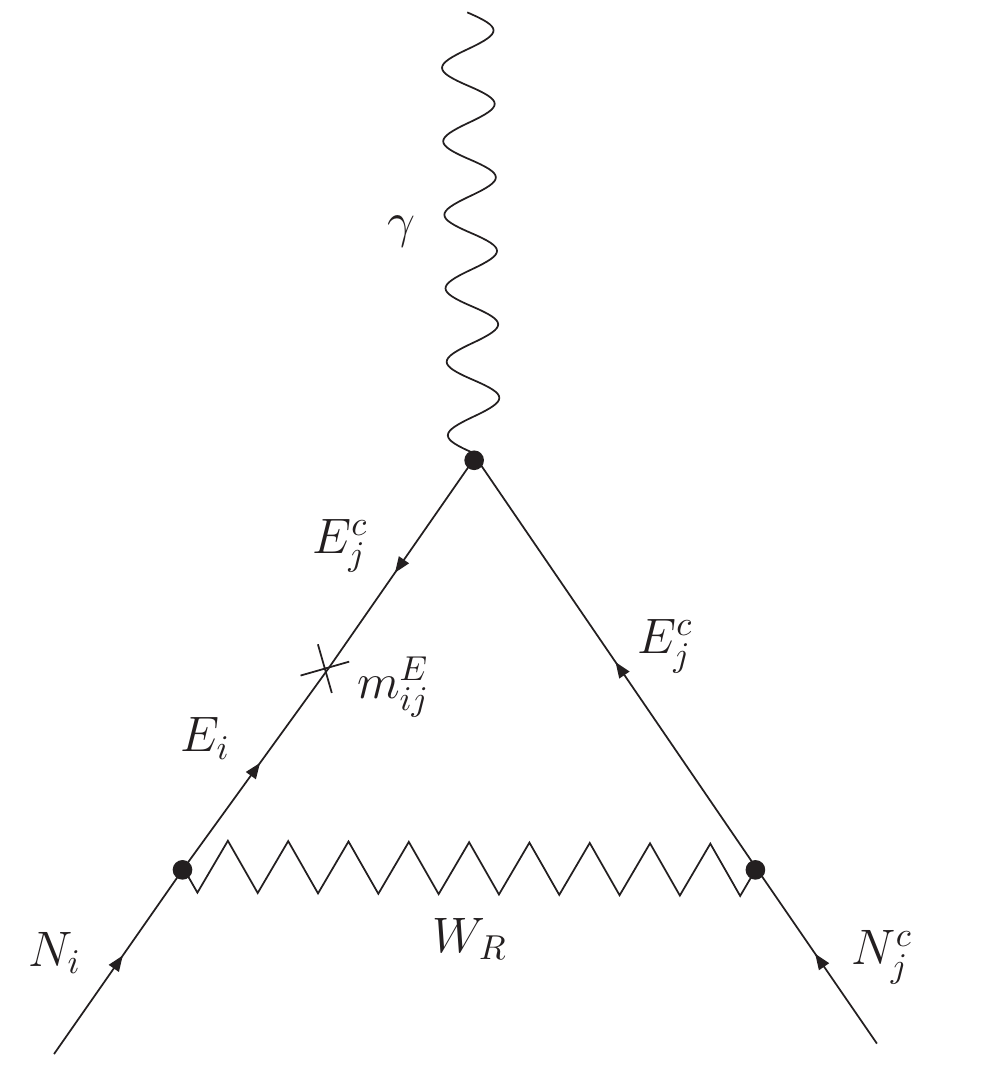}
\includegraphics[scale=0.7]{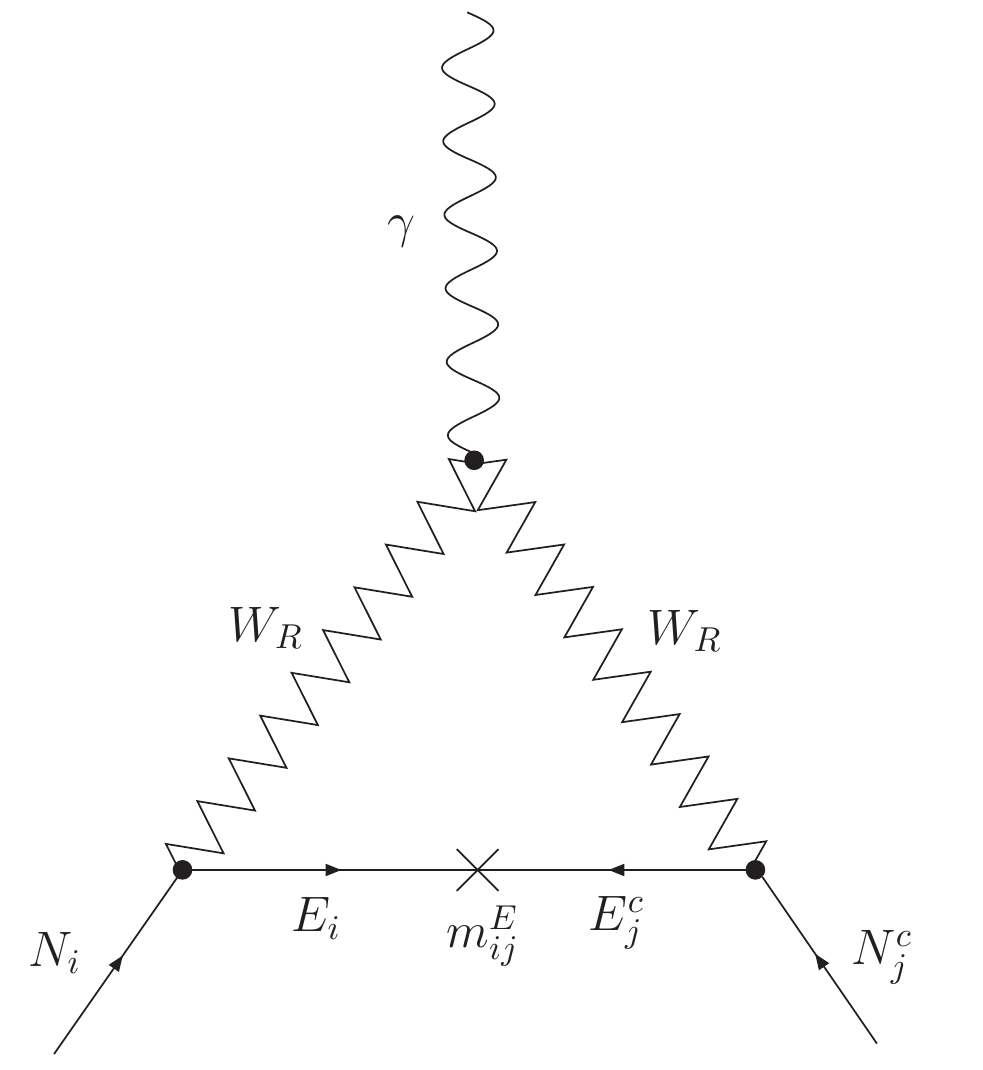}
\end{center}
\caption{The one-loop diagram generating an electromagnetic transition between $N_i$ and $N_j^c$ in the left-right model discussed in the text.}
\label{Fig:EM.transitions}
\end{figure}

Let us assume a left-right model where the sterile neutrinos, $\{N_h, N_h^c \}$ and $\{N_{h^\prime}, N_{h^\prime}^c\}$ are accommodated in $SU(2)_R$ doublets together with charged leptons, 
\begin{align}
L_h		& = \begin{pmatrix} N_h \\ E_h^- \end{pmatrix}					&
L_h^c		& = \begin{pmatrix} E_h^{c \, +} \\ N_h^{c} \end{pmatrix}			&
L_{h^\prime}	& = \begin{pmatrix} N_{h^\prime} \\ E_{h^\prime}^- \end{pmatrix}			&
L_{h^\prime}^c	& = \begin{pmatrix} E_{h^\prime}^{c \, +} \\ N_{h^\prime}^{c} \end{pmatrix}
\label{Eq:sterile.neutrino.SU2R.doublets}.
\end{align}
It is straightforward to define a framework where the breaking of $SU(2)_R$ gives large masses of order $v_R$ to the charged components of these doublets while the sterile neutrinos remain light; that is, it is possible to get $m_h = 50\text{ MeV}$, $m_{h'} = 10\text{ MeV}$, while $m_E \approx 500\text{ GeV}$.  The mass matrix of the charged leptons, $m^E_{ij} E_i E^c_i$ ($i, j \in \{h, h^\prime\}$), is then what gives rise to the decay of $\nu_h$ along with the magnetic dipole moment of $\nu_{h^\prime}$:~these \abbr{EM} transitions will be generated by the processes shown in \fig{EM.transitions}.
In particular, the $\nu_h \to \nu_{h^\prime} \gamma$ decay is induced by $i = h$, $j = h^\prime$ while the magnetic dipole moment is gotten for $i = j = h^\prime$.
To reproduce the scenario discussed in \cite{Masip:2012ke}, it is necessary that $m^E_{ij}$ obey the hierarchy $m^E_{h h^\prime} \ll m^E_{h^\prime h} \approx 0.5 m^E_{h h} \ll m^E_{h^\prime h^\prime}$.  For the magnetic dipole moment $\mu_{h^\prime}$ of $\nu_{h^\prime}$ we obtain
\begin{equation}
\mu_{h'} = \frac{e g_R}{4 \pi^2 v_R} \, r \, \frac{r^4 + 12 r^2 - 13 + \inp{6r^4 - 20 r^2}  \log r^2}{8 \inp{r^2 - 1}^3} \,,
\label{Eq:magnetic.moment.muhprime}
\end{equation}
with $r = m_{E}/M_R$ and taking the limit $m_{h^\prime} \ll m_E, M_R$. Values around $r = 1.5$, $v_R= 500\text{ GeV}$, and $g_R = 0.8$ can yield $\mu_{h^\prime} \approx 10^{-5}\text{  GeV}^{-1} = 0.3\E{-6}\mu_B$.  While this magnetic moment is relatively large for a neutral particle, it is not large enough to explain the number of events at \abbr{DAMA}/\abbr{LIBRA} without a significant increase in the flux (recall that \eq{N.Events} states that the number of events is proportional to $\mu^2 \Phi$).  Unfortunately, this magnetic moment seems close to the maximum value that a loop of exotic charged particles can generate without appealing to either strong dynamics or higher multiplicities.  As such, another means of generating large magnetic moments will be required if the flux is to remain  a fraction of that of $\nu_\mu$.

Before considering the enhancement of the dipole moment, it is useful to estimate the impact of this setup on \abbr{DM} experiments.  We will assume that the total number of atmospheric heavy neutrinos, $\nu_{h^\prime}$, is $0.5$\% that of muon neutrinos, and will take the muon neutrino abundance (in \fig{flux.nu.and.recoils}) from \cite{Honda:2006qj}.  For the energy distribution we will take
\begin{equation}
\Phi_{h^\prime}(E) \propto
	\begin{cases}
	E		& m_{h^\prime} \le E < 40\text{ MeV}	\\
	E^{-2.7}  	& E \ge 40\text{ MeV}
	\end{cases}
\label{Eq:flux.nuhprime}
\end{equation}
We will set $m_{h^\prime} = 10\text{ MeV}$
and consider a lifetime long enough so that we can ignore $\nu_{h^\prime}$ decays.

\begin{figure}[!bt]
\begin{center}
\includegraphics[width=0.48\linewidth]{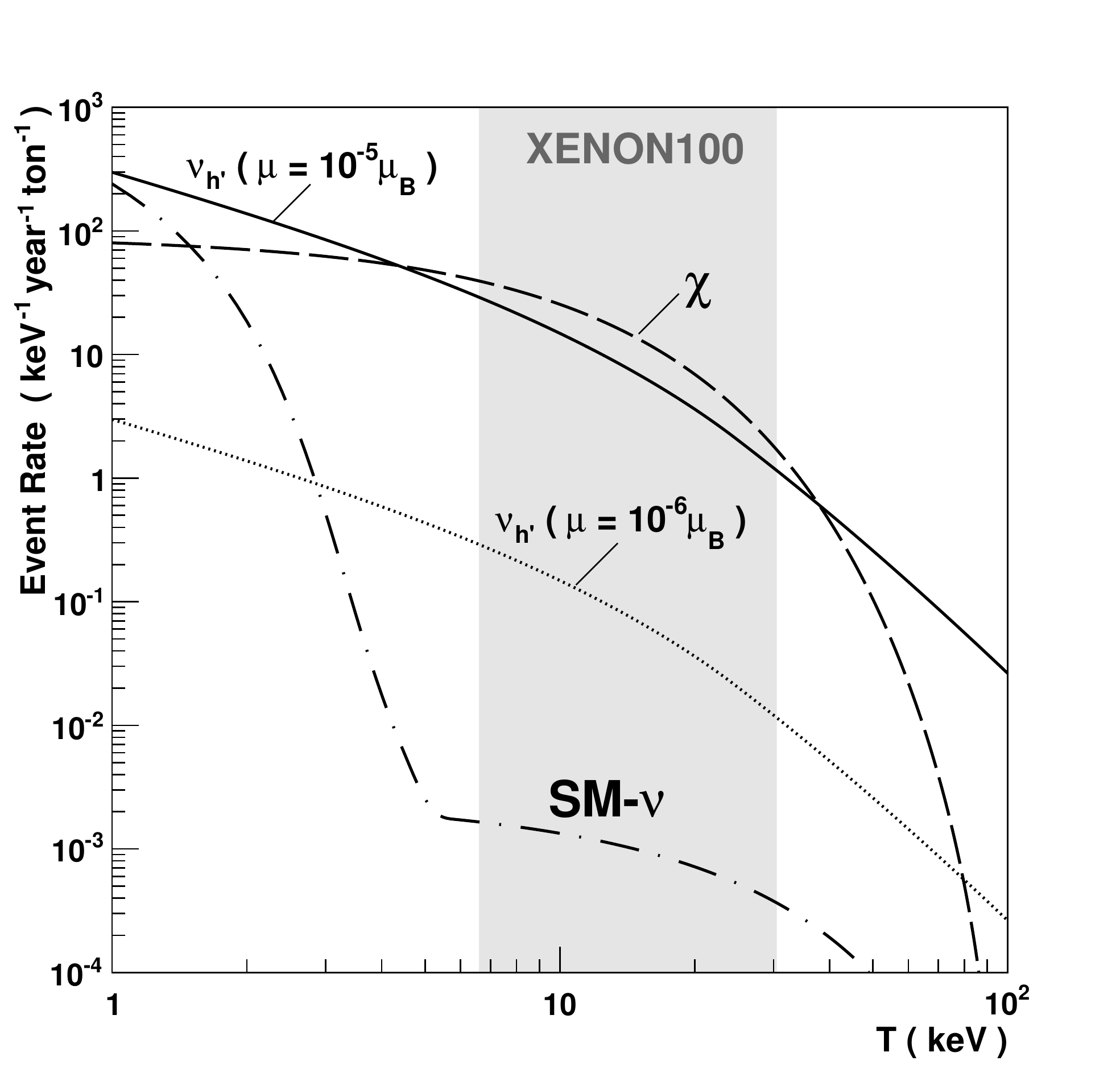}
\includegraphics[width=0.48\linewidth]{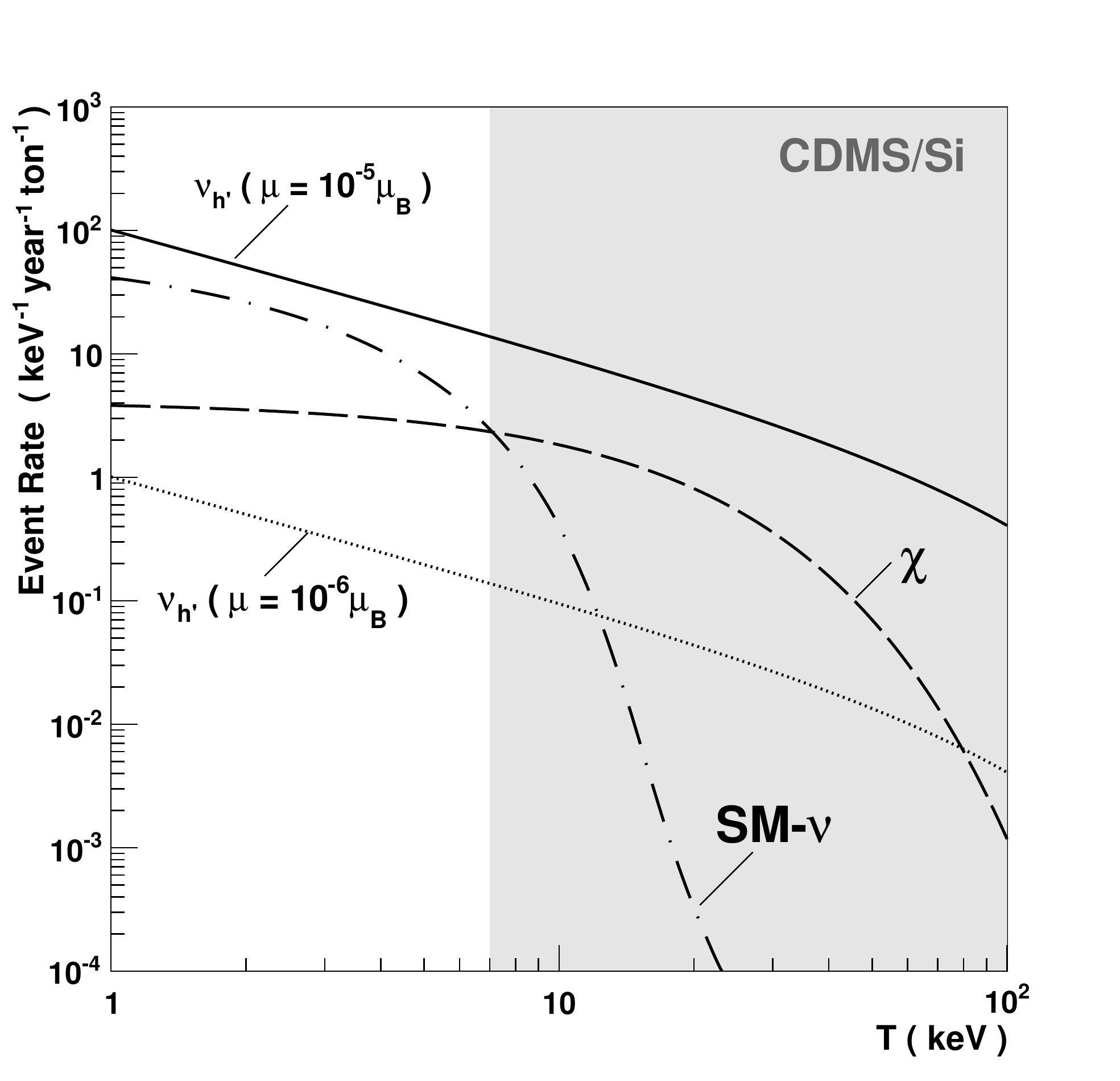} 
\end{center}
\caption{Recoil distribution of $\nu_{h^\prime}$ events for $\mu = 10^{-6} \mu_B$ (dotted) and $\mu = 10^{-5} \mu_B$ (solid) at XENON100 (left) and \abbr{CDMS}/Si (right). Also shown are the number of events from standard neutrino interactions (dot-dashed) as well as the current XENON100 limit\cite{Aprile:2012nq} from a $55$ GeV \abbr{WIMP} with a cross section $\sigma_{\chi N} = 2\E{-45} \text{ cm}^2$ (dashed).  The shaded regions show the acceptance regions of the experiments: XENON100 has $T \in [6,30]$ keV; \abbr{CDMS}/Si has $T \in [7,100]$ keV.}
\label{Fig:events.XENON100.and.CDMS}
 \end{figure}

With the above criteria, the expected number of $\nu_{h^\prime}$ events per year and ton at XENON100 and \abbr{CDMS}/Si is shown in \fig{events.XENON100.and.CDMS} for $\mu_{h^\prime} = 10^{-6} \mu_B$ (dotted) and $\mu_{h^\prime} = 10^{-5} \mu_B$ (solid).
For comparison, we also include the number of events from standard neutrinos along with the current XENON100 limit\cite{Aprile:2012nq} of a $55$ GeV \abbr{WIMP} having a cross section of $\sigma_{\chi N} = 2\E{-45}\text{ cm}^2$.  The figure demonstrates that although the number of $\nu_{h^\prime}$ events that we obtain is larger than the background from ambient neutrinos, an observable signal at both XENON100 and \abbr{CDMS}/Si would require magnetic moments around $10^{-5} \mu_B$ rather than the $10^{-6} \mu_B$ generated in the left-right symmetric model (specifically, for $\mu_{h^\prime} = 10^{-5} \mu_B$ we estimate $171$ events per year and ton at XENON100 and $208$ at \abbr{CDMS}/Si). 
\begin{figure}[!th]
\begin{center}
\includegraphics[scale=0.7]{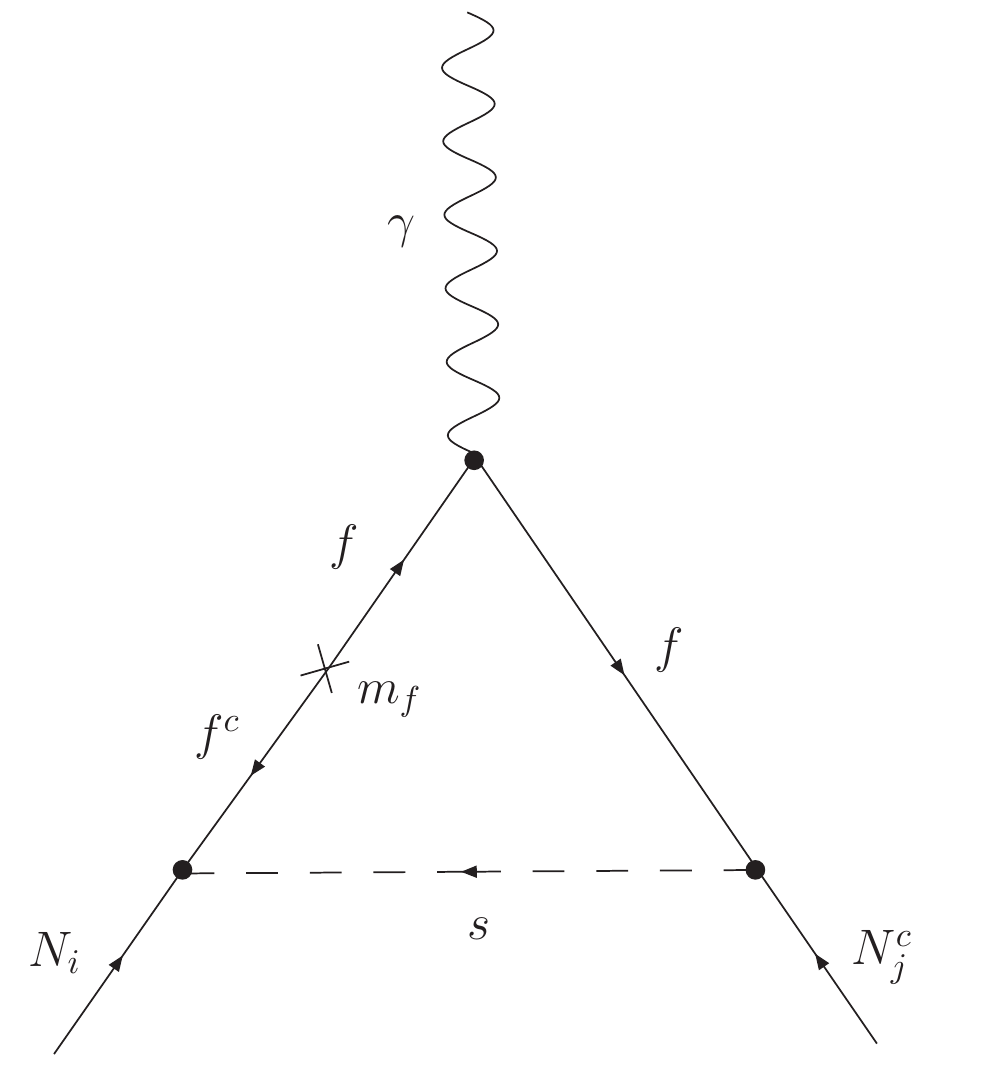}
\includegraphics[scale=0.7]{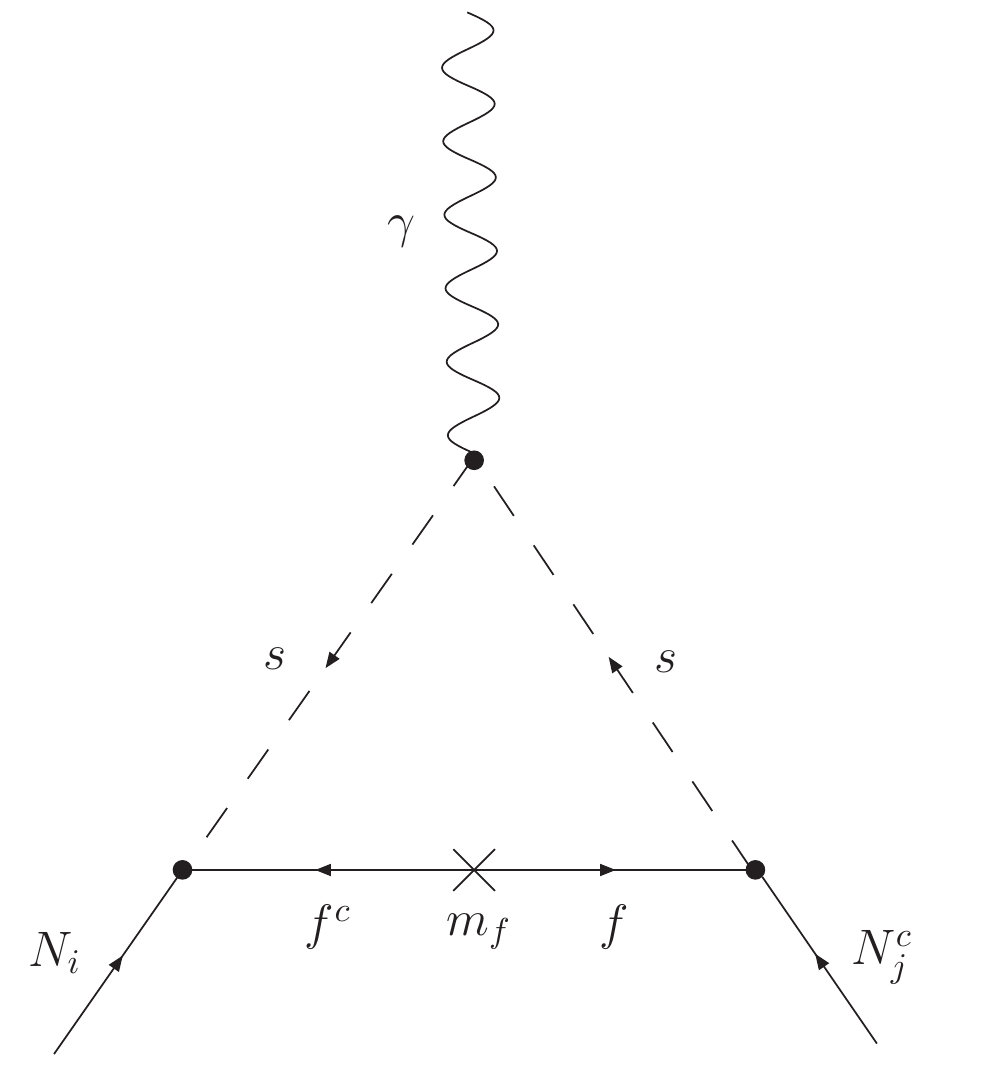}
\end{center}
\caption{Some one-loop diagrams generating a magnetic dipole moment for the $\nu_{h^\prime}$ (composed of the two-component spinors $N$ and $N^c$) from the millicharged fermions $f$ and $f^c$ and the scalar $s$.  The charges are $Q_f = - Q_{f^c} = Q_s = \epsilon e$ which permits the Yukawa couplings $f^c s N$ and $f s^\dagger N^c$.}
\label{Fig:magnetic.dipole.moment.from.millicharges}
\end{figure}
Even if one could obtain a magnetic dipole moment of $10^{-5} \mu_B$, it would still be too small to have an impact on \abbr{DAMA}/\abbr{LIBRA} and \abbr{CoGeNT}. 

To realize the scenario for \abbr{DAMA}/\abbr{LIBRA}, \abbr{CoGeNT}, \abbr{CDMS}/Si, and XENON100 that we have described in the previous section, we need $\mu_{h^\prime} \sim 10^{-2} \mu_B$ and also a much steeper flux, \textit{i.e.}, a mechanism that reduces the dipole energy on its way from the atmosphere to the detector and yields a spectral index $\gamma\approx 10$ deduced in \eq{Flux.E.dependence}.  Remarkably, both effects may be correlated: a much larger value of $\mu_{h^\prime}$ could make the neutrino interact several times in the rock shielding the detector and, at the same time, would allow for the strong signal observed at \abbr{DAMA}/\abbr{LIBRA} and \abbr{CoGeNT}.  More specifically, we estimate the energy loss through ionization of a dipole relative to that of a muon as
\begin{equation}
- \left. \deriv{E}{x} \right|_{N} \approx - \frac{ \mu^2 m_e^2 }{ \pi \alpha } (\beta \gamma)^2 \left. \deriv{E}{x} \right|_{\mu}
\label{Eq:stopping.power.relative.muon}
\end{equation}
with $\beta \gamma = p/m$ and $m_e$ the mass of the electron. For $\mu_{h^\prime} =10^{-6}\mu_B$ the stopping power of $1$ km of rock would be negligible, while for values around $10^{-2}\mu_B$ the dipole would lose a significant fraction of energy.  This latter value of $\mu_{h^\prime}$ would also be compatible with the \abbr{DAMA}/\abbr{LIBRA} and \abbr{CoGeNT} signals for a total dipole flux smaller than the atmospheric neutrino flux (again, the number of events is proportional to $\Phi_{h'} \times \mu_{h^\prime}^2$). Although a complete calculation of the $\nu_{h^\prime}$ flux at the detector site is beyond the scope of this work, we find interesting that both conditions are simultaneously satisfied for that value of $\mu_{h^\prime}$.

While a dipole moment of $10^{-2} \mu_B$ can explain the dark matter experiments, it also opens up the possibility of events at neutrino experiments measuring electron recoils.  Employing \eq{Def.Tmax} with $M = m_e$ yields that the maximum energy transfered to an electron is $T_\text{max} = 0.6$ MeV for an $m = 10$ MeV, $E = 13$ MeV dipole.  Such small electron recoils mean that $\nu_{h^\prime}$ would not be visible at \abbr{MiniBooNE} (where the anomaly is explained by the decays $\nu_{h}\to \nu_{h^\prime} \gamma$ of the heavier neutrino $\nu_h$), since the resulting electron energies would be well below the energy cuts.  The atmospheric flux of $\nu_{h^\prime}$ would also be invisible in solar neutrino experiments like Super Kamiokande for the same reason: the energy threshold there is around $4$ MeV. The remaining possibility of detection is then sub-MeV solar neutrino experiments like Borexino.  We estimate that the  atmospheric $\nu_{h^\prime}$ flux is around $10^{-10}$ that of the solar neutrino flux and that  the ionization cross section of $\nu_{h^\prime}$ is about $10^7$ times larger than the quasielastic $\sigma(\nu_e A \to e A)$; therefore, the number of $\nu_{h^\prime}$ events at Borexino should be negligible.

Of course, we are still facing the fact that the required magnetic moments are much larger than the ones expected in natural completions of the model. We think, however, that such values are possible if, for example, our sterile neutrino $\nu_{h^\prime}$ is coupled to exotic particles of low mass ($10$--$100$ MeV) having electric millicharge.  These particles should annihilate before primordial nucleosynthesis \cite{Davidson:1993sj}, and the massless paraphoton that they imply could give an interesting contribution to the dark radiation suggested by some experiments \cite{Hou:2012xq,Hinshaw:2012aka,Ade:2013zuv,Sievers:2013ica}.  For example, one Dirac fermion $\{ f, f^c \}$ plus one scalar $s$ both millicharged and with sizeable Yukawa couplings to $\{N_{h^\prime}, N_{h^\prime}^c\}$ could induce the large magnetic moment that is required through the diagrams in \fig{magnetic.dipole.moment.from.millicharges}.  We estimate that these diagrams produce
\begin{equation}
\mu \sim \frac{1}{16\pi^2} \frac{y^2 \epsilon}{m_{f}} \sim 10^{-2} \mu_B \pfrac{\epsilon}{10^{-3}} \pfrac{10 \text{ MeV}}{m_f}
\label{Eq:magnetic.dipole.moment.from.millicharges.estimate}
\end{equation}
which is clearly around the desired magnetic moment.

\section{Summary and discussion}
\label{Sec:conclusion}

We have shown how an atmospheric dipole may reconcile the seemingly inconsistent data of direct-search experiments, and we have provided a concrete model as evidence of its viability.  
We hope that the additional data of current and forthcoming experiments (ANAIS \cite{Amare:2012ex}, COUPP \cite{Behnke:2011zz}, LUX \cite{Fiorucci:2013yw}, DEAP \cite{Boulay:2012hq}, etc.)~will settle whether the current signals are due to \abbr{DM}, to new physics of atmospheric or solar \cite{Harnik:2012ni,Pospelov:2012gm} origin, or to some unidentified background.

\section*{Acknowledgments}

This work has been partially supported by MICINN of Spain (FPA2010-16802, FPA2010-17915, FPA2012-39489.C04.04 and Consolider-Ingenio {\bf Multidark} CSD2009-00064 and {\bf CPAN} CSD2007-00042), and by Junta de Andaluc\'{\i}a
(FQM 101,330,3048).

\bibliography{%
	\bibpath/neutrinos,%
	\bibpath/exp-neutrinos,%
	\bibpath/exp-muons,%
	\bibpath/exp-dark_matter,%
	\bibpath/exp-cosmology,%
	\bibpath/dark_matter-detecting,%
	\bibpath/detectors-dark_matter,%
	\bibpath/microcharges,%
	\bibpath/dark_matter%
}

\end{document}

%% file: cmds/cite_commands.tex
\newcommand{\eq}[1]{Eq.~\eqref{Eq:#1}}
\newcommand{\eqn}[1]{\eqref{Eq:#1}}
\newcommand{\fig}[1]{Figure~\ref{Fig:#1}}

\newcommand{\Sec}[1]{Section~\ref{Sec:#1}}
\newcommand{\tbl}[1]{Table~\ref{Table:#1}}

%% file: cmds/math-calculus.tex
		\newcommand{\pderiv}[3][]{%
			\ifthenelse{\equal{#1}{}}%
				{%
				\frac{ \partial #2}{ \partial #3 }%
				}%
				{%
				\frac{ \partial^{#1} #2}{ \partial {#3}^{#1} }%
				}%
			}
		\newcommand{\deriv}[3][]{%
			\ifthenelse{\equal{#1}{}}%
				{%
				\frac{ d #2}{ d #3 }%
				}%
				{%
				\frac{ d^{#1} #2}{ d {#3}^{#1} }%
				}%
			}
		\newcommand{\fderiv}[3][]{%
			\ifthenelse{\equal{#1}{}}%
				{%
				\frac{ \delta #2}{ \delta {#3} }%
				}%
				{%
				\frac{ \delta^{#1} #2}{ \delta {#3}^{#1} }%
				}%
			}
		

%% file: cmds/math-algebra.tex
		\newcommand{\abs}[1]{ \mathopen{}\left| {#1}\right| }


%% file: cmds/math-fractions.tex
		\newcommand{\half}{\frac{1}{2}}			

%% file: cmds/math-parentheses.tex
		\newcommand{\inp}[2][0cm]{\mathopen{}\left(#2\parbox[h][#1]{0cm}{}\right)}
		
		\newcommand{\pfrac}[2]{\mathopen{}\left(\frac{#1}{#2}\right)}
		

%% file: cmds/format-science-misc.tex
	\newcommand{\E}[1]{\times 10^{#1}}
	